\documentclass[a4paper,fleqn,usenatbib]{mnras}
\usepackage[T1]{fontenc}
\usepackage[english]{babel}
\usepackage{ae,aecompl}
\usepackage{graphicx}	% Including figure files
\usepackage{amsmath}	% Advanced maths commands
\usepackage{amssymb}	% Extra maths symbols
\usepackage{multicol}        % Multi-column entries in tables
\usepackage{bm}		% Bold maths symbols, including upright Greek
\usepackage{pdflscape}	% Landscape pages
\usepackage{caption}
\usepackage{adjustbox}
\usepackage{natbib}
\usepackage{graphicx}	% Including figure files 
\usepackage{amsmath}	% Advanced maths commands
\usepackage{amssymb}	% Extra maths symbols
\usepackage{xcolor}
\title[The VMC Survey - XXXV. Model fitting of LMC Cepheid light curves.]{The VMC Survey - XXXV. Model fitting of LMC Cepheid light curves}

\author[Fabio Ragosta et al.]{Fabio Ragosta,$^{1,2,3}$
	Marcella Marconi,$^{3}$
	Roberto Molinaro,$^{3}$\thanks{E-mail: roberto.molinaro@inaf.it}
	Vincenzo Ripepi,$^{3}$
	\and  
	Maria Rosa L. Cioni,$^{4}$ 
	Maria Ida Moretti,$^{3}$
	Martin A.T. Groenewegen,$^{5}$
	\and Samyaday Choudhury,$^{6,7}$
	Richard de Grijs,$^{6,7,8}$
	Jacco Th. van Loon,$^{9}$
	\and Joana M. Oliveira,$^{9}$
	Valentin D. Ivanov,$^{10}$
	Carlos Gonzalez-Fernandez$^{11}$	
	\\
	$^{1}$Dipartimento di Fisica, Via Cinthia, I-80126 Fuorigrotta, Naples, Italy\\
    $^{2}$INFN sez. di Napoli Compl. Univ. di Monte S. Angelo, Edificio G, Via Cinthia, I-80126 - Napoli, Italy \\
	$^{3}$INAF-Osservatorio Astronomico di Capodimonte, via Moiariello 16, I-80131, Naples, Italy\\
	$^{4}$Leibniz-Institut f\"{u}r Astrophysik Potsdam, An der Sternwarte 16, D-14482 Potsdam, Germany \\
	$^{5}$Koninklijke Sterrenwacht van Belgi\"e, Ringlaan 3, B--1180 Brussels, Belgium  \\
	$^{6}$Department of Physics and Astronomy, Macquarie University, Balaclava Road, Sydney, NSW 2109, Australia\\
	$^{7}$Research Centre for Astronomy, Astrophysics and Astrophotonics, Macquarie University, Balaclava Road, Sydney, NSW 2109, Australia\\
	$^{8}$International Space Science Institute--Beijing, 1 Nanertiao, Zhongguancun, Hai Dian District, Beijing 100190, PR China\\
	$^{9}$Lennard-Jones Laboratories, School of Chemical and Physical Sciences, Keele University, ST5 5BG, UK\\
	$^{10}$European Southern Observatory, Karl-Schwarzschild-Str. 2, D-85748 Garching bei M\"{u}nchen, Germany \\
	$^{11}$Institute of Astronomy, University of Cambridge, Madingley Road, Cambridge CB3 0HA, UK\\
}

\date{Accepted XXX. Received YYY; in original form ZZZ}

\pubyear{2019}

\begin{document}
	\maketitle
	
	% Abstract of the paper
	\begin{abstract}
		We present the results of the light curve model fitting technique applied to optical and near-infrared photometric data for a sample of 18 Classical Cepheids (11 fundamentals and 7 first overtones) in the Large Magellanic Cloud (LMC). We use  optical photometry from the OGLE III database and  near--infrared photometry obtained by the  European Southern Observatory public survey ``VISTA  near--infrared survey of the Magellanic Clouds system''. Iso--periodic nonlinear convective model sequences have been computed for each selected Cepheid in order to  reproduce the multi--filter light curve amplitudes and shape details. 
         The inferred individual distances provide an intrinsic weighted mean value for the LMC distance modulus of $\mu_0=18.56$ mag with a standard deviation of 0.13 mag.
          We derive also the Period--Radius, the Period--Luminosity and the Period--Wesenheit relations that are  consistent with similar relations in the  literature. 
           The intrinsic masses and luminosities of the best--fitting models show that all the investigated pulsators are brighter than the predictions of the canonical evolutionary mass--luminosity relation,  suggesting a significant efficiency of non--canonical   phenomena, such as overshooting, mass loss and/or rotation.
			\end{abstract}
	% Select between one and six entries from the list of approved keywords.
	% Don't make up new ones.
	
	\section{Introduction}
	Classical Cepheids (CCs) are a class of pulsating stars widely
        used to calibrate the extragalactic distance scale, through
        secondary distance indicators.  
		Their role as tool to measure distance is based on the
        relation they show between the period of pulsation and their intrinsic
        luminosity, known as Period--Luminosity (PL) relation.
        From an evolutionary point of
        view, a CC is a star with an intermediate mass  (from $~ 3 $M$_{\odot}$ to $ 13 $M$_{\odot}$), in the
        central helium burning phase (covering an age range from $\sim10~$Myr to $\sim200~$Myr, see \citealt{and17}), 
        that passes through the instability strip as it evolves bluewards and then redwards
        (blue loop excursion), at roughly constant luminosity for each
        given mass.  From a theoretical point of view,  CCs, like
          all the other classes of pulsating stars, obey to a relation
          between the oscillation period and their mean density, as
          demonstrated by \citet{edd26}. By combining this relation
          with the Stefan--Boltzmann law one obtains a
          Period--Luminosity--Mass--Temperature relation. For CCs, this
          relation can be reduced to a Period--Luminosity--Color
           (PLC) relation,  because the theory of stellar evolution predicts the 
        existence of a mass--luminosity relation (MLR), whose
        coefficients depend on
        the assumed metal and helium abundances.  Currently adopted PL relation can be seen as a
        projection  of the PLC relation  on the PL plane
        \citep[and references therein]{fre91,bon99, cap05}. In other
        words, the PL relation shows an intrinsic dispersion related to
        the finite width of the instability strip. Obviously, for each
        filter combination, both the PLC and its projection
        on the PL plane critically depend on the MLR.  Since the
        efficiency of noncanonical phenomena (such as core
        overshooting, mass loss, rotation) significantly affects the
        MLR, in order to determine precise distances using CCs, a detailed theoretical evaluation of the impact of these
        processes needs to be assessed.
	
	Several authors have discussed the
        effect of mass loss and core overshooting \citep[see][and references therein]{chi93, bon99, cap05, kel06, nei12, mar13}, as well as of rotation
        \citep{and16} on CC properties. These studies are often
        related to the so called mass discrepancy problem, first
        outlined by \citet{sto69} and \citet{chr70} and subsequently
        confirmed by additional investigations. According to
        these studies, the CC evolutionary mass (inferred from the
        comparison between theoretical isochrones and observations in
        the color--magnitude diagram) was found to be systematically higher than
        the ``pulsational'' one based on the Period--Mass--Radius
        relation\footnote{From the combination of the Period--density relation and the Stefan--Boltzman law it is also possible to obtain a Period--Mass--Radius relation, which is useful to estimate the masses of Cepheids if their radii are known, and vice versa. According to the linear adiabatic theory the pulsation period of variables is related to mass and radius through the equation  $P =\alpha (M/M_\odot)^\beta (M/M_\odot)^\gamma$ \citep{fri72}, which can be linearized easily in logarithmic space  ($\log P = \log \alpha + \beta \log (M/M_\odot) + \gamma \log (R/R_\odot)$) thus obtaining the PMR relation.} \citep{fri72, bon01} or other methods
        based on the theory of pulsation \citep{bon02, cap05}.  
        \citet{kel06} and \cite{mar13} adopted the
        model fitting of multi--filter light, radial velocity and radius curves
        to address the mass discrepancy. This is done  
        through direct comparison of the observed and predicted variations along a
        pulsation cycle, the latter based on nonlinear convective
        pulsational models \citep[see][for a detailed discussion of  the method]{bon00, bon02, mar13}.  
	
        The ``VISTA near--infrared $\textit{Y},\textit{ J},\textit{ K}_\mathrm{s}$ survey of the Magellanic
        Clouds system'' \citep[VMC -][]{cio11} covers the Magellanic system with
        deep Near--Infrared (NIR) ($\textit{Y}$,$\textit{ J}$, $\textit{K}_\mathrm{s}$ filters) VIRCAM \citep[VISTA InfraRed Camera; ][]{dal06} photometry using the ESO/VISTA telescope
       \citep{eme06}. The main science goals of the VMC are the
        study of the spatially--resolved star--formation history (SFH)
        and the determination of the 3D structure of the whole Magellanic
        system. Particularly useful for the latter aim are pulsating variables
        such as RR Lyrae stars and CCs that have been  the subject of several
        studies in the context of the VMC survey, as distance indicators and
        stellar population tracers \citep[see e.g.][]{rip12a, rip12b, rip14, rip15, rip16, rip17, mor14, mor16, mur15, mur18}.  

        \citet{mar17} presented the model fitting of
        multi--wavelength light curves and, when available, radial
        velocity curves of 12 Small Magellanic Cloud (SMC) CCs
        whose NIR observations were secured as part of the VMC data.
        The inferred stellar parameters and
        individual distances permitted to constrain not only the mean
        distance modulus of the SMC but also the behaviour of the
        investigated stars in the MLR, PL, Period--Radius (PR) and
        Period--Wesenheit (PW) relations.\footnote{The Wesenheit
        	magnitudes include a color term  with a
                  coefficient that corresponds to the ratio between total
        	to selective extinction in the selected filter pair \citep{mad82, cap00}, thus
        	making the Wesenheit relations reddening free.}
        
        In this paper we extend this to a sample of $11$ fundamental (F) and $7$ first
        overtone (FO) CCs in the Large Magellanic
        Cloud (LMC), that are within the field of view of the VMC survey.
        
As regards the organization of the paper, in Section 2 we
        discuss the sample selection,  in Section 3 we describe the adopted model fitting technique.
        The application of this technique to the selected LMC CCs and the implications of our results for the MLR, the PR, the PL and PW relations are described in sections 4 and 5 respetively.
        The final section includes a summary and  perspectives.

\section{Selection of the sample}
The selected sample of CCs is composed of $11$ and $7$ F and FO pulsators, respectively, that cover a range in oscillation period from $\sim1$ to $\sim30$ days.  The selected CCs have
optical photometry from the OGLE III database\footnote{When this work began OGLE IV data were not available. We checked for possible changes between the two data releases finding that the number of 	points in the \textit{V} and \textit{I} bands does not increase by more than a few 	percents for most sources in our sample. Only  for three
	stars the photometric observations increase significantly, but this does not affect the results of our 	method,  because a good phase coverage of light curves was already available in OGLE III.}
\citep{sos10} and NIR photometry from the VMC database \citep[see e.g.][for a description of VMC light curves]{cio11, rip16, rip17}.  In particular we used aperture photometry data from the
Cambridge Astronomy Survey Unit (CASU) and the Vista Science Archive (VSA)\citep[see][for details]{cro12, gon18}. The sample is selected in order to span a wide range in period,
luminosity and shape of the light curves from the OGLE III database. The period values spanned by our sample do not include the range of the so called 'bump' Cepheids (8-12 days). However, this period range has already been analyzed, using the same pulsating code as in the current work, in \citet{bon02}, where the authors selected a couple of LMC Cepheids, one with the bump on the rising branch and the other on the decreasing branch.

Altough the number of the selected CCs does not represent the entire LMC, they let us test the prediction capabilities of the model fitting technique in a stellar system.
We note that obtaining a statistically significant extension of the selected target number would be extremely time consuming to reach convergence of our hydrodynamical pulsation code.

\begin{figure}
	\centering
	\includegraphics[trim=20 140 0 20,clip,scale=.45]{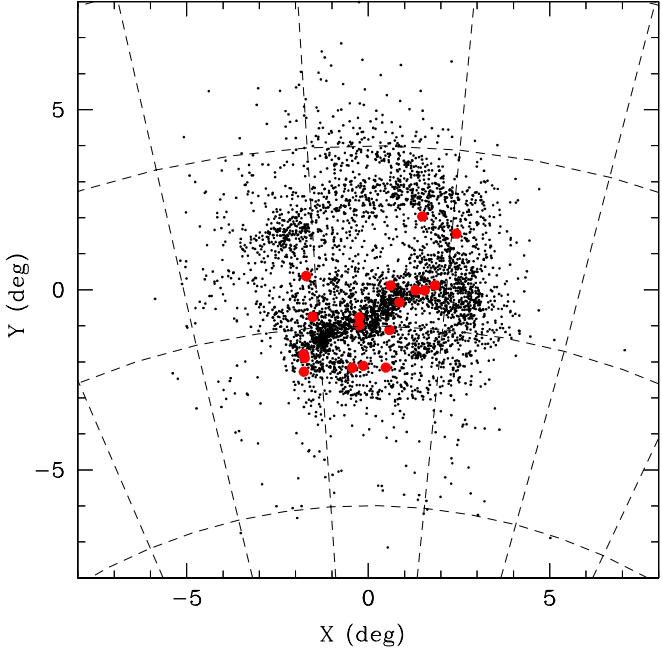}
	\caption{Distribution on the sky of the CCs investigated in this paper (red filled circles).  For comparison the whole sample of known CCs in the LMC from the
		OGLE survey (black dots) is shown.  X and Y are defined as in \citet{van01} with $\alpha_0=81^\circ$ and $\delta_0= -69^\circ$ }
	\label{plot}
\end{figure}

The  distribution in right ascension and declination of the selected CCs is shown in Fig.\ref{plot}, where all known CCs in the LMC from the OGLE survey \citep[][]{sos15} are shown for comparison. The identification, the period, and the mean visual  magnitude, of the selected CCs are listed in the first three columns of Table~\ref{tab1}; these values were taken from \citet[][]{sos15}. As shown in Table~\ref{tab1} the selected CC sample encompasses wide ranges of periods and mean magnitude thus allowing us to check the predictive capabilities of the model fitting technique over a large interval of CC observed properties. 
%\onecolumn

\section{The model fitting technique}
The fitting technique adopted to find the best model reproducing the observations, is similar to that described in  \citet{mar17}:
\begin{itemize}
	\item  both observed and modeled photometric curves are phased in order to have the maximum of light  at the same phase in a given reference band: e.g. in this work the maximum of light in the \textit{V} band is at phase 0.
	\item for each modeled photometric band, we estimated the shifts in magnitude ($\delta \mu$) and phase ($\delta \phi$) that provide
	the best match between modeled and observed light curves. Specifically, these two parameters have been obtained by minimizing the following  $\chi^2$ equation:
\end{itemize}

\begin{equation}\label{chi}
\chi^2=\frac{1}{N_{\textrm{bands}}}\sum_{j}^{\text{N}_{\textrm{bands}}}\frac{1}{N_{\textrm{DOF}}^j}\sum_{i}^{\text{N}^j_{\textrm{pts}}}\frac{[\text{m}^j_i-(\textit{M}^j_{\textrm{mod}}(\phi_i^j+\delta \phi^j)+\delta \mu^j)]^2}{\sigma_i^j}
\end{equation}

where the two indices $j$ and $i$ run respectively on the number of bands, $N_{\textrm{bands}}$, and on the number of epochs, $N_{\textrm{pts}}$, the observed phases, magnitudes and errors are indicated with $\phi_i^j$ and $\text{m}_i^j$, $\sigma_i^j$ respectively, while  $M^j$ is the absolute magnitude of the pulsating model evaluated at the same phase of observations plus the shift $\delta\phi$. To evaluate the model at a given phase the theoretical light curves have been interpolated using a smooth spline. In the above formula, the term $N_{DOF}^j=N^j_{\textrm{pts}}-2$ is  the number of degrees of freedom for the j--th band. We note that, because the initial phasing procedure, described above, is the same for observations and models, the fitted values of $\delta \phi$ are typically small ($\sim 0.02$) and represent just fine tuning phase shift values. On the other hand, the fitted parameter $\delta\mu^j$ represents the apparent distance modulus in the $j^{\text{th}}$ band.

Following the same approach as in \citet{mar17}, for each
selected target CC, we	built isoperiodic model
sequences at fixed mass, varying the effective temperature and
in turn the luminosity level.   This allowed us  to obtain
a sample  of light curve models with different shapes,
magnitudes and periods. Using this sample of models, we
selected the model  best matching 	the observed
curves. The isoperiodic sequences are built using 	the
typical elemental composition of the LMC
\citep[$\textit{Y}$=$0.25$; $\textit{Z}$=$0.008$; see][for
details]{rom08, luc98}. Once we found the best fitting
$T_{\text{e}}$, we 	built another sample of models varying
the mass but fixing  the	obtained best fitting 	temperature, again selecting the best fitting model matching the observed curves. Thus we were able to evaluate the
mass, the luminosity, the effective temperature of the star and in turn its individual apparent distance modulus in each  selected band.

\onecolumn
\begin{figure}
	\centering
	\includegraphics[scale=0.55]{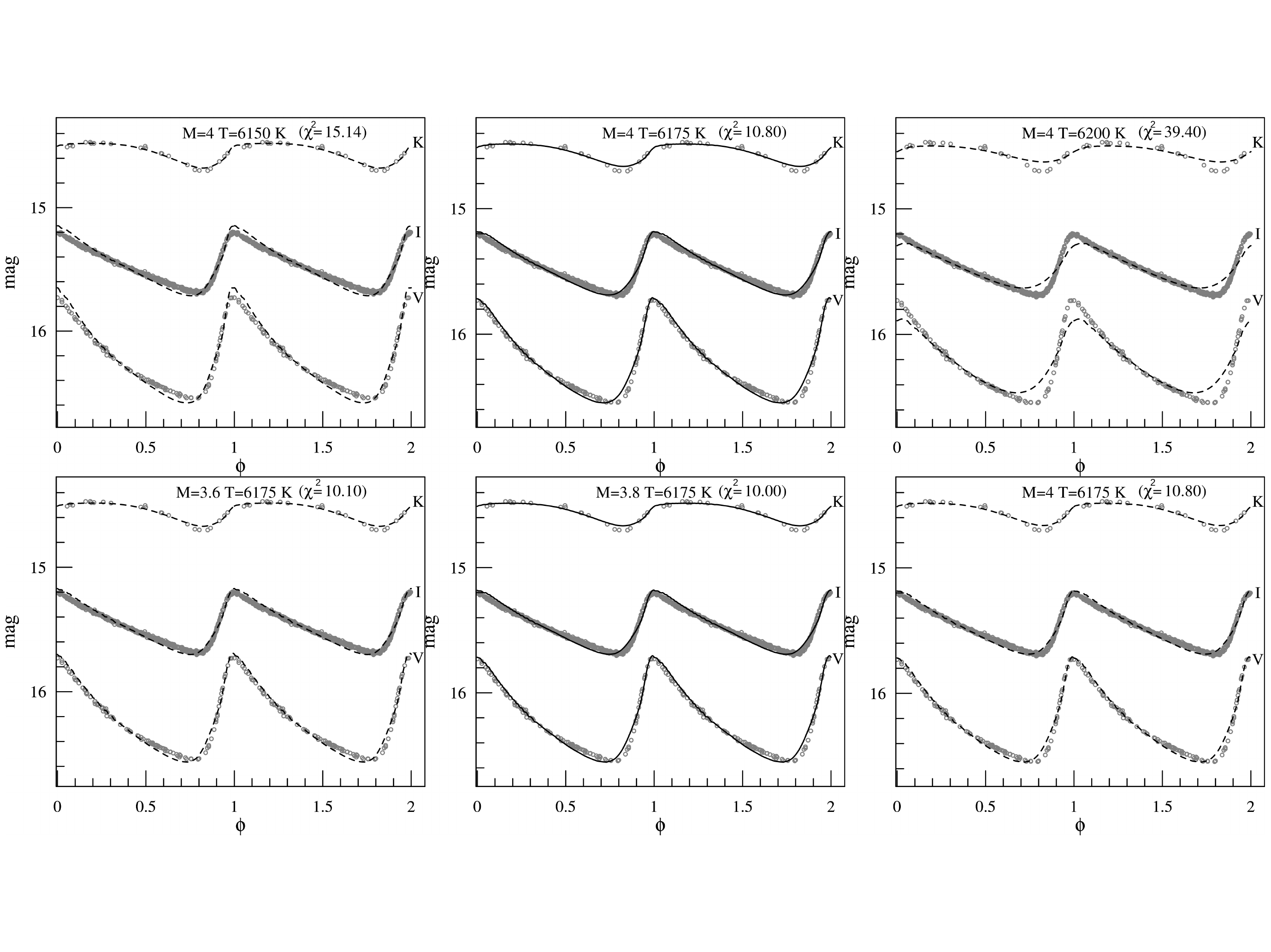}
	\captionof{figure}{Comparison of the observed $\textit{V}$, $\textit{I}$ and $\textit{K}_s$ light curves of the variable OGLE LMC CEP 2138 and the fitted theoretical light curves. Data are plotted with grey symbols, while models are plotted with lines. In the top panels we show the models calculated assuming a fixed mass, namely ($M=4.0$M$_{\odot}$) and varying effective temperature. Once the model with best $T_e$ (namely $T_e=6175 \textrm{K}$) is found,  (solid line), the effective
		temperature is maintained fixed at its best  value and the $\chi^2$ in equation \ref{chi} is minimized by varying the mass. Models with varying mass and fixed $T_e$ are showed in the bottom panels, where the final best fitting model is again indicated with a solid line and is characterized by $T_e=6175 \textrm{K}$ and $M=3.8$M$_\odot$. The $\chi^2$ values of the fit are also labeled in each panel.}\label{TempMassGrid}. 
\end{figure}
\twocolumn
\begin{figure}
	\centering
	%	\begin{minipage}{\columnwidth}
	\includegraphics[trim=10 10 200 450,clip,scale=0.6]{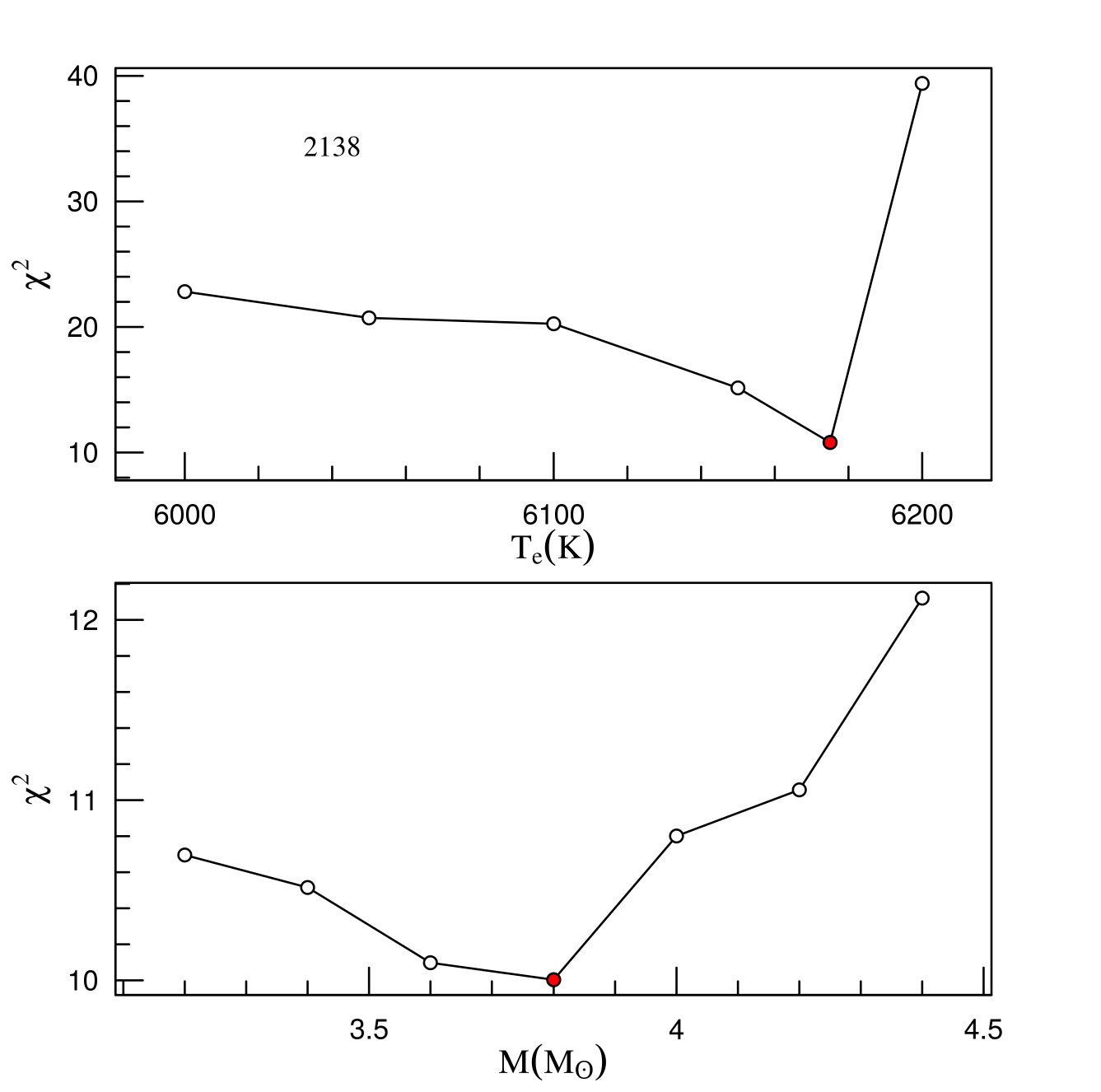}
	\caption{Total reduced $\chi^2$  values obtained from the fitting procedure applied 	to  $OGLE\_LMC\_CEP\_2138$ are shown as a function of the 	model effective temperature (top) and 	mass (bottom). The values of the best effective temperature and mass are indicated by the red dots.}
	\label{masstemp}
	%\end{minipage}
\end{figure}

In Fig.~\ref{TempMassGrid} we show an example of the model fitting dependence on the assumed effective temperature (fixed mass $ M=4.0$M$_{\odot}$) and stellar mass (fixed $T_{\textrm{eff}}=6175 \textrm{K}$). The $\chi^2$ analysis for models with fixed mass
identifies as best fit effective temperature $T_{\textrm{eff}}=6175 \textrm{K}$, as shown in the top panel of the quoted figure. Then varying the mass (Fig.~\ref{TempMassGrid} bottom panels), one obtains the best fitting value $M=3.8$M$_{\odot}$.  Since it is difficult to evaluate by--eye the quality of the fit shown in Fig.~\ref{TempMassGrid}, in Fig.~$\ref{masstemp}$ we show the total reduced $\chi^2$ as a function of the model  effective temperature (top) and  mass (bottom).

\begin{figure}
	\includegraphics[trim=10 10 200 450,clip,scale=0.6]{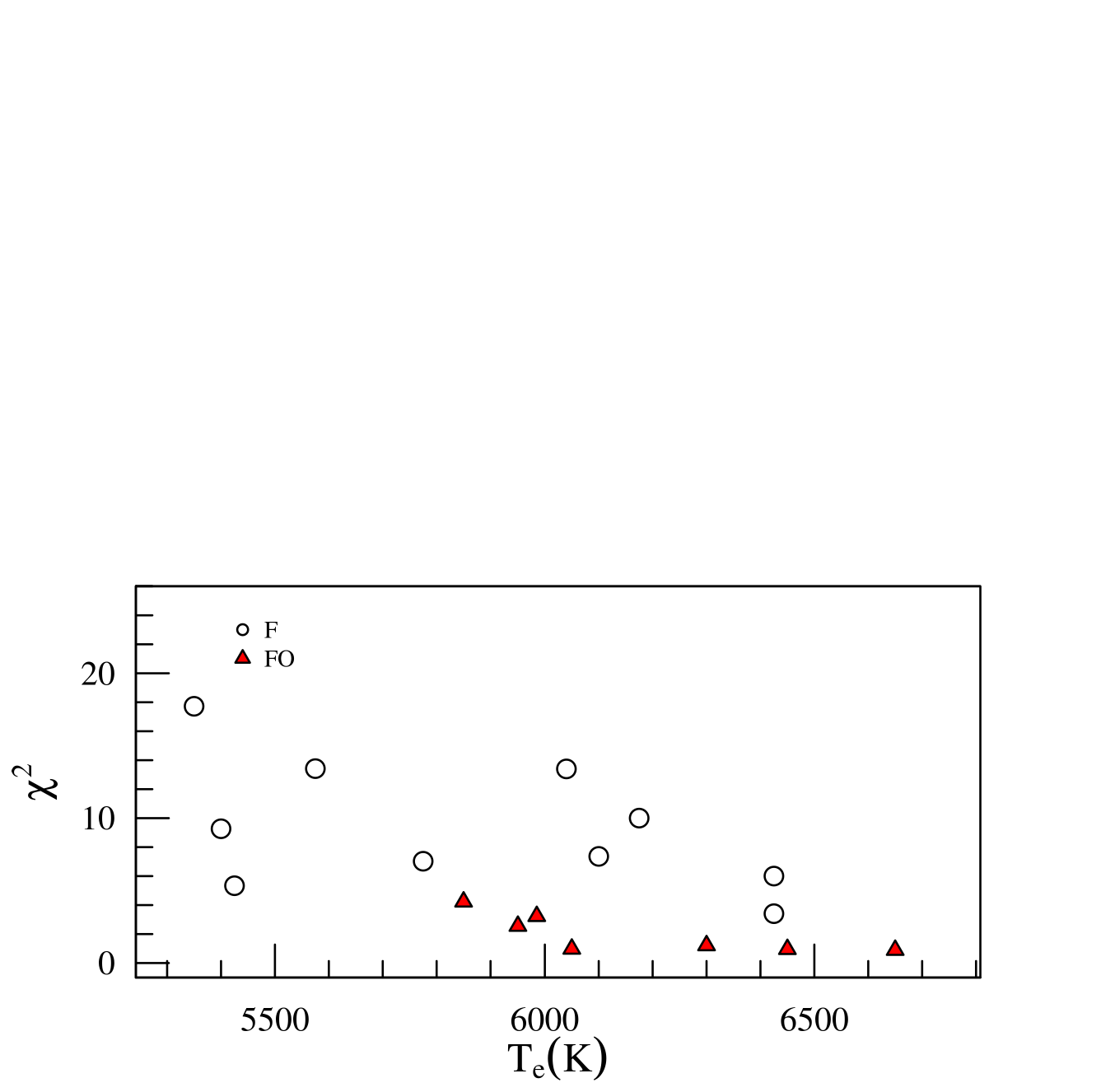}
	%\centering
	\caption{The  $\chi^2$ values obtained from the fitting procedure and  listed in Table~\ref{tab1} are plotted against the best values of effective temperature for all sources of the selected sample: F pulsators are shown using empty circles, while FOs are plotted with red triangles. }
	\label{ffig-chi2VsT}
\end{figure}

\begin{figure}
	\includegraphics[scale=1.5]{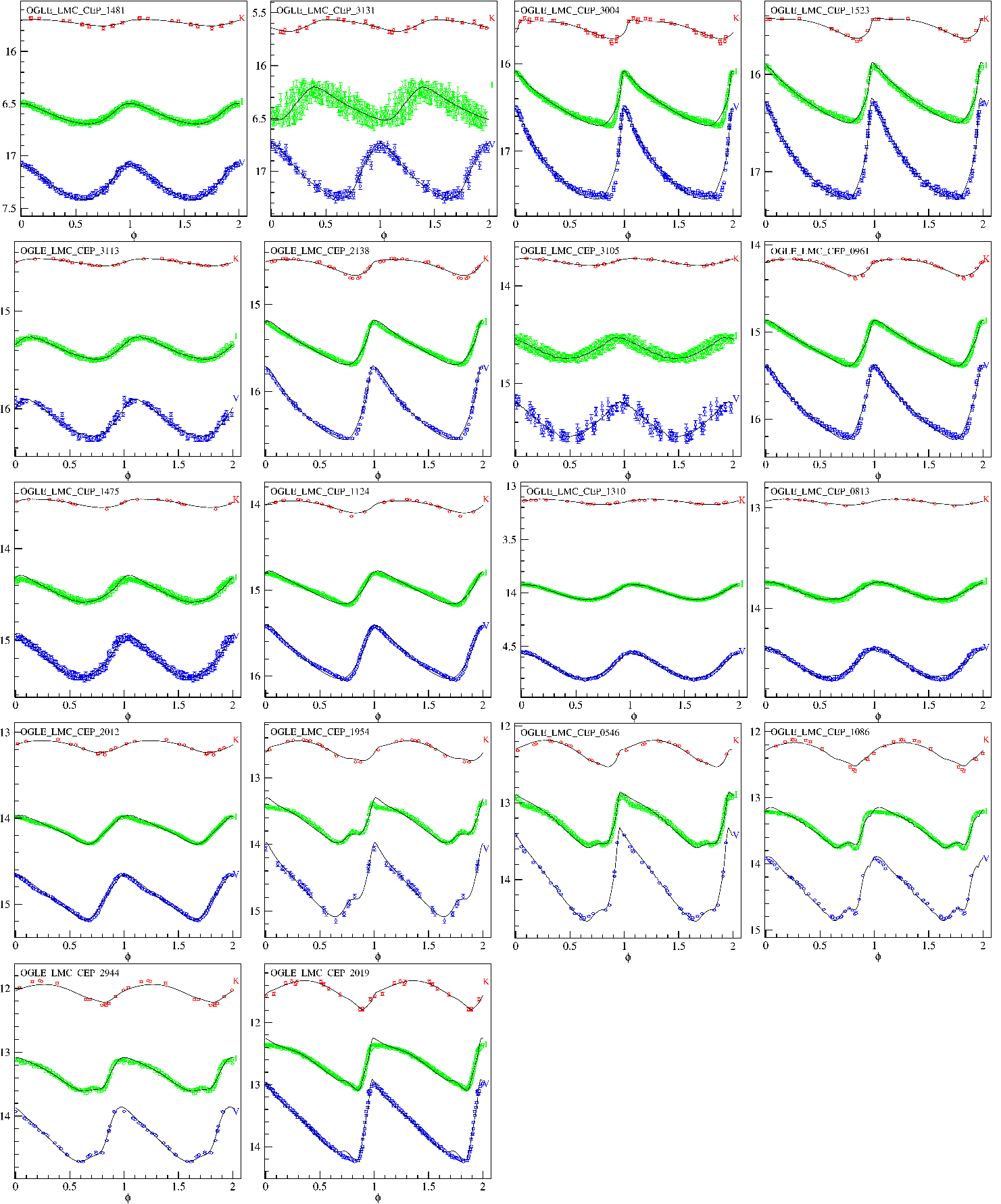}
	\centering
	\caption{Best fitting models for all selected CCs compared with the observed light curves.  Black lines represent the theoretical models, while the $\textit{V}$, $\textit{I}$  and $\textit{K}_\mathrm{s}$  observed light curves are labeled together with star identifications.}
	\label{fig-bestFit}
\end{figure}

Looking at the best $\chi^2$ value in the figure and
also at those reported in the Tab.~\ref{tab1} it is evident
that they are not always close to the expected canonical value
$\chi^2=1$. This is due both to error underestimation of the
observations  and to the difficulty to reproduce exactly
light curves with more complex shapes (see also
Fig.~\ref{fig-bestFit}). The presence of features in the light
curves of pulsating variables, which make them more complex from the shape point of view, is due to the  coupling between pulsation
and convection which becomes more important moving towards
the red boundary of the instability strip
\citep[see][and references therein]{bms99}.
On this basis, we  expect a correlation between the $\chi^2$ values
and the best fitting effective temperature in the sense that lower
$\chi^2$ values correspond to higher effective temperatures.  This
trend is evident in  Fig.~\ref{ffig-chi2VsT}  where the $\chi^2$
values of Table~\ref{tab1} are plotted against the best fitting
effective temperature $T_e$\footnote{Note that the source OGLE\_LMC\_CEP\_0546  does not appear in Fig.~\ref{ffig-chi2VsT} because of its $\chi^2$ value is out of the y-axis range.},  with lager $\chi^2$ values populating
the zone of lower effective temperaures. Moreover a clear separation
can be seen between F and FO models, the latter having  smaller
$\chi^2$ values and as well as known higher effective temperatures.

\onecolumn

\begin{table}
	\centering
	\caption{Properties of the target CCs and of the associated best--fitting models. From left to right: OGLE identification, observed period, mean $\textit{V}$ magnitude and pulsation mode  \citep[from the OGLE III database][]{sos10}, best--fitting model mass, luminosity, effective temperature, mixing length 	parameter, inferred distance modulus in the	$\textit{V}$, $\textit{I}$, and $\textit{K}_\mathrm{s}$ bands and associated errors, absolute distance modulus with  associated error, magnitude correction to refer distance modulus to LMC barycenter, reddening with associated error, mean radius and $\chi^2$ value. For all CCs the
		error in the inferred mass and temperature is $\Delta T= 25$ K and $\Delta M= 0.2$M$_{\odot}$, respectively; the assumed composition is $\textit{Y}=0.25, Z= 0.008$, except for $OGLE\_LMC\_CEP\_2019$ for which $\textit{Y}=0.30$ (see text).}
			\begin{adjustbox}{max width=1.3\textwidth}
		% latex table generated in R 3.6.1 by xtable 1.8-4 package
% Wed Jul 24 12:33:07 2019
\begin{tabular}{ccccccccccccccccccccc}
  \hline
%ID OGLE & $\textit{P}$ & $\langle V \rangle$ & Mode & $M$ &$log(L/L_\odot)$ &$T_{\textrm{eff}}$ & $\alpha_{ML}$ & $\mu_V \pm \sigma_{\mu_V}$ &  $\mu_I \pm \sigma_{\mu_I}$ &  $\mu_K \pm \sigma_{\mu_K}$ &  $\mu_0\pm\sigma_{\mu_0}$ &  $E(B-V)\pm\sigma_{E(B-V)}$ & $R$ & $\chi^2$ & $\mu_0^{corr}$\\ 
ID & P & $\langle V\rangle$ & Mode & $M$ & $log(L/L_\odot)$ & $T_{eff}$ & $\alpha_{ML}$ & $\mu_V$ &  $\mu_I$ &  $\mu_K$ &  $\mu_0$ &  $\mu_0^{corr}$ &  $E(B-V)$ & $R$ & $\chi^2$ \\ 
  & (days) & (mag) & & $($M$_\odot)$ & (dex) & (K) & & (mag) &  (mag) &  (mag) & (mag) & (mag) & (mag) &$($R$_\odot)$ & \\ \hline
1481 & 0.922 & 17.28 & FO & 3.00 & 2.62 & 6650 & 1.50 & 19.14$\pm$0.10 & 18.89$\pm$0.07 & 18.49$\pm$0.08 & 18.42$\pm$0.09 & 0.002 & 0.24$\pm$0.05 & 15.4 & 0.9 \\ 
  3131 & 1.095 & 17.01 & FO & 2.80 & 2.63 & 6450 & 1.80 & 18.89$\pm$0.05 & 18.72$\pm$0.03 & 18.52$\pm$0.02 & 18.47$\pm$0.02 & 0.02& 0.14$\pm$0.01 & 16.6 & 1.0 \\ 
  3004 & 1.524 & 17.15 & F & 3.00 & 2.65 & 6425 & 1.60 & 19.08$\pm$0.08 & 18.86$\pm$0.04 & 18.55$\pm$0.01 & 18.48 $\pm$0.01 & -0.04& 0.19$\pm$0.03 & 17.2 & 6.0  \\ 
  1523 & 1.572 & 16.88 & F & 2.80 & 2.64 & 6425 & 1.60 & 18.78$\pm$0.08 & 18.62$\pm$0.04 & 18.44$\pm$0.02 & 18.39$\pm$0.02 & 0.05 & 0.13$\pm$0.02 & 17.0 & 3.4  \\ 
  3113 & 2.068 & 16.08 & FO & 5.20 & 3.15 & 6300 & 1.60 & 19.27$\pm$0.05 & 19.06$\pm$0.03 & 18.77$\pm$0.01 & 18.71$\pm$0.01 & 0.008& 0.18$\pm$0.02 & 31.5 & 1.2 \\ 
  2138 & 3.011 & 16.21 & F & 3.80 & 3.03 & 6175 & 1.70 & 19.04$\pm$0.07 & 18.85$\pm$0.05 & 18.60$\pm$0.05 & 18.54$\pm$0.05 & 0.04& 0.16$\pm$0.02 & 28.7 & 10.0  \\ 
  3105 & 3.514 & 15.38 & FO & 4.80 & 3.33 & 6050 & 1.60 & 18.96$\pm$0.06 & 18.82$\pm$0.04 & 18.62$\pm$0.03 & 18.58$\pm$0.03 & 0.011& 0.13$\pm$0.01 & 42.2 & 1.0  \\ 
  0961 & 3.711 & 15.87 & F & 3.90 & 3.11 & 6100 & 1.70 & 18.90$\pm$0.07 & 18.76$\pm$0.06 & 18.54$\pm$0.07 & 18.49$\pm$0.07 & 0.018& 0.13$\pm$0.02 & 32.5 & 7.4 \\ 
  1475 & 4.387 & 15.19 & FO & 5.60 & 3.48 & 5985 & 1.51 & 19.14$\pm$0.09 & 19.01$\pm$0.08 & 18.77$\pm$0.06 & 18.73$\pm$0.06 & 0.03& 0.13$\pm$0.01 & 51.2 & 3.2  \\ 
  1124 & 4.457 & 15.81 & F & 5.00 & 3.29 & 6040 & 1.70 & 19.24$\pm$0.08 & 19.06$\pm$0.06 & 18.79$\pm$0.03 & 18.73$\pm$0.03 & 0.014& 0.16$\pm$0.02 & 40.5 & 13.4  \\ 
  1310 & 5.126 & 17.28 & FO & 5.70 & 3.54 & 5950 & 1.49 & 18.78$\pm$0.04 & 18.71$\pm$0.05 & 18.60$\pm$0.05 & 18.58$\pm$0.06 & 0.015& 0.07$\pm$0.01 & 55.5 & 2.6  \\ 
  0813 & 5.914 & 14.54 & FO & 7.00 & 3.66 & 5850 & 1.53 & 18.93$\pm$0.07 & 18.87$\pm$0.05 & 18.75$\pm$0.03 & 18.73$\pm$0.03 & 0.02& 0.07$\pm$0.01 & 66.2 & 4.2  \\ 
  2012 & 7.458 & 14.95 & F & 6.50 & 3.54 & 5775 & 1.90 & 18.98$\pm$0.03 & 18.88$\pm$0.02 & 18.71$\pm$0.01 & 18.68$\pm$0.01 & 0.008& 0.10$\pm$0.01 & 59.4 & 7.0  \\ 
  1954 & 12.950 & 14.61 & F & 5.30 & 3.69 & 5575 & 1.90 & 19.00$\pm$0.07 & 18.81$\pm$0.05 & 18.61$\pm$0.02 & 18.55$\pm$0.02 & 0.04& 0.14$\pm$0.02 & 75.8 & 13.4  \\ 
  0546 & 15.215 & 14.03 & F & 5.20 & 3.77 & 5575 & 1.70 & 18.59$\pm$0.11 & 18.58$\pm$0.11 & 18.57$\pm$0.10 & 18.56$\pm$0.11 & -0.003& 0.01$\pm$0.02 & 83.9 & 44.2  \\ 
  1086 & 17.201 & 14.35 & F & 5.40 & 3.75 & 5350 & 1.90 & 18.86$\pm$0.09 & 18.76$\pm$0.08 & 18.64$\pm$0.08 & 18.61$\pm$0.08 & -0.03& 0.08$\pm$0.02 & 89.5 & 17.7  \\ 
  2944 & 20.320 & 14.30 & F & 6.90 & 3.95 & 5400 & 1.84 & 19.30$\pm$0.06 & 19.15$\pm$0.06 & 18.83$\pm$0.07 & 18.78$\pm$0.08 & -0.010& 0.17$\pm$0.02 & 108.6 & 9.3  \\ 
  2019 & 28.103 & 13.64 & F & 7.70 & 4.11 & 5425 & 1.70 & 19.00$\pm$0.07 & 18.83$\pm$0.06 & 18.67$\pm$0.04 & 18.62$\pm$0.05 & 0.013& 0.12$\pm$0.02 & 131.9 & 5.3  \\ 
   \hline
\end{tabular}
	
	\end{adjustbox}\label{tab1}
\end{table}

\twocolumn

\section{Application to the selected variables}
The procedure detailed in the previous section was applied to all CCs in Table~\ref{tab1} and the corresponding models  are shown in Fig.~\ref{fig-bestFit}.

We note that for the longest period CC in our sample (namely $OGLE\_LMC\_CEP\_2019$) we had to vary also the elemental composition in order to reproduce the observed light curves. 
For this star we adopted $\textit{Y}=0.30$,
$\textit{Z}=0.008$. The inferred intrinsic stellar parameters,
namely  the effective temperature, the luminosity and the mass
of the  best fitting models are reported in Table~\ref{tab1}
with their errors. 

In particular, the errors on the parameters obtained from the fitting procedure are estimated as the difference between the best values and the parameters  of the closest models to the best fitting one on the Mass--Temperature grid. 
As regard the error on the mass and  temperature of the best model, we have adopted the steps of the  Mass--Temperature model grid (0.2 M$_{\odot}$ and 25 K respectively) generated for our analysis.

The quoted table contains also the  unreddened distance moduli $\mu_0$ and the $\textit{E(B--V)}$ values for all stars considered in the present work. They have been calculated by fitting the Cardelli law \citep{car89} to the inferred apparent distance moduli in the $\textit{V}$, $\textit{I}$, $\textit{K}_\mathrm{s}$ bands.

A simple statistical analysis  of the values reported in
Tab.~\ref{tab1} gives a mean value of the reddening equal to
$\textit{E(B--V)}=0.13$ mag with a standard deviation of 0.05 mag, while the
inferred mean distance  modulus for the LMC is equal to
$\mu_0=18.59$ mag with a standard deviation equal to 0.12
mag. Weighting the fitted  parameters with their errors provides almost the same values, with mean distance modulus that is equal to $\mu^{\textrm{wt}}_0=18.63$ mag ($\sigma^{\textrm{wt}}=0.10$ mag) and the mean reddening $\textit{E(B--V)}^{\textrm{wt}}=0.11$ mag ($\sigma^{\textrm{wt}}=0.04$ mag). 

In order to take into account the effect of the inclination of the plane of the LMC with respect to the sky on the barycentric distance estimation, we have also calculated the magnitude corrections (see Table~\ref{tab1}) for every CC of our sample according to the geometric model by \citet{van01}. A statistical analysis of the distances obtained by including the quoted corrections does not change the results reported above about the LMC distance.

As stated in the previous section, the $\chi^2$ values exhibit a large scatter (see Table~\ref{tab1}) indicating that the more complex light curve shapes are modeled with lower accuracy and larger residuals. Therefore, we decided also to weigh the best parameters using the $\chi^2$ values to define the weights ($\textrm{wts}=1/\chi^2$) in order to favour those models that better describe the observed light curve shapes. The resulting weighted distance modulus is equal to $\mu^{\textrm{wt}_{\chi^2}}_0=18.56$ mag with a standard deviation $\sigma^{\textrm{wt}_{\chi^2}}=0.13$ mag. We assume that this value is our best estimate of the LMC distance.  Using the same $\chi^2$--weighted statistics for the excess, we obtain a mean value equal to $\textit{E(B--V)}^{\textrm{wt}_{\chi^2}}=0.15$ mag with a standard deviation equal to $\sigma^{\textrm{wt}_{\chi^2}}=0.05$ mag.

The quoted errors represent only the statistical uncertainities,
while the systematic is difficult to estimate but depends on the
physical and numerical assumptions of the current model sets
as well as on residual uncertainties of the adopted atmosphere
models. Moreover, the above results for the distance modulus do not
take into account projection effects, related to the fact  that
CCs are not located at the centre of the LMC.

\section{Results}
In this section we use the results obtained for the intrinsic stellar parameters of the investigated CCs to determine constraints both on the predicted MLR and PR relations as well as on the PL and PW relations, at least for the assumed
elemental composition. 
\begin{figure}
	\includegraphics[trim=10 10 200 530,clip,scale=0.7]{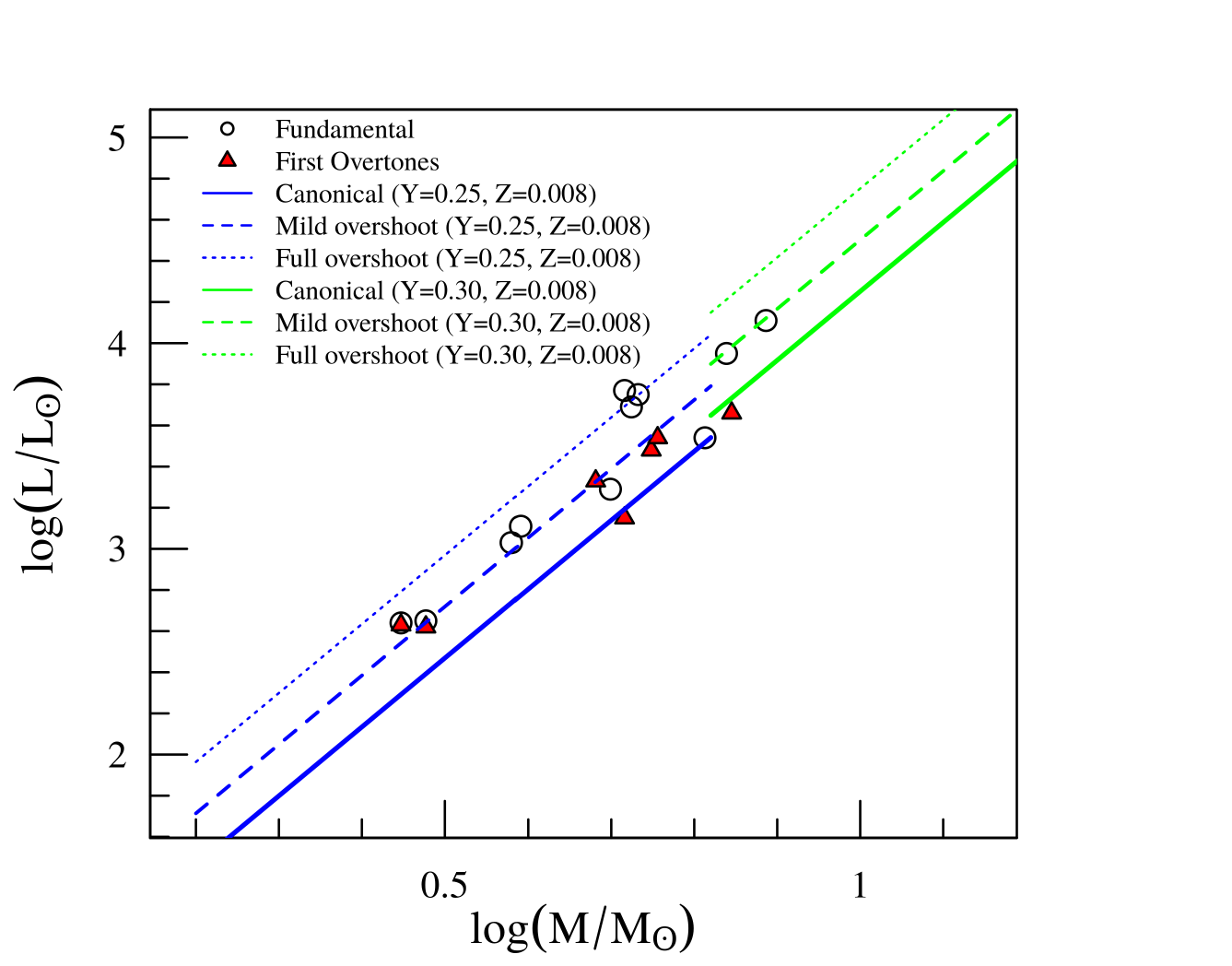}
	\captionof{figure}{Predicted MLR based on the model fitting results for both F (empty circles) and FO CCs (red filled triangles). The best fitting model location in the MLR plane is compared with an evolutionary MLR obtained by neglecting 
		mass loss, core overshooting and rotation (labelled ''Canonical") and with the relations obtained by assuming mild or full overshooting (see  text for detail).}\label{MLrelation}		
\end{figure}
%\onecolumn
\begin{figure}
	\centering
	\includegraphics[trim=10 10 200 450,clip,scale=0.7]{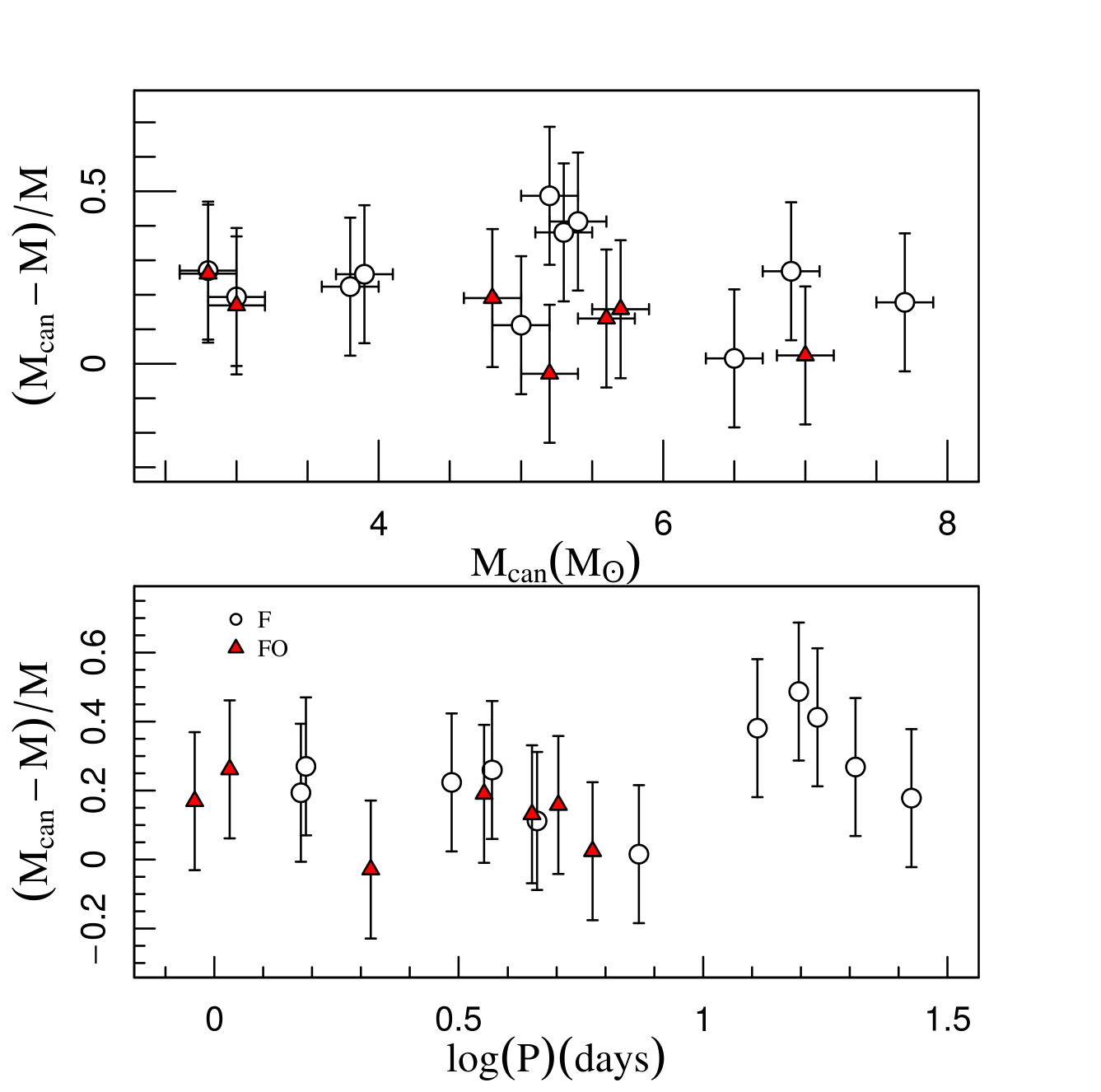}
	\captionof{figure}{ Deviation of the best fitting stellar mass from the value corresponding  to the canonical mass for both F (empty circles) and FO (red triangles) CCs.}
	\label{MLoss}
\end{figure}
\twocolumn
\subsection{The Mass--Luminosity Relation}
Figure~\ref{MLrelation}  displays the MLR for the investigated CCs whose intrinsic stellar parameters were derived from the best fitting models listed in Table~\ref{tab1}. These data are compared with the predicted canonical (no overshooting, no mass loss)
MLR  \citep{bon00} (solid lines) and with the relations obtained by  increasing the zero point of the canonical MLR by $0.25$ dex (dashed lines) and $0.5$ dex (dotted lines) to reproduce the effect of mild and full overshooting\footnote{Corresponding to an extension of the
	extra--mixing region beyond the Schwarzschild border of about $0.2H_p$ where $H_p$ is the pressure scale height \citep{chi93}.}, respectively \citep[see][for details]{chi93, bon99}. Inclusion of mass loss and/or  rotation would produce a similar increase in the Cepheid luminosity level at fixed mass \citep[see][for details]{nei12}.
As the light curves of $OGLE\_CEP\_LMC\_2019$ are best reproduced
adopting a different value of the helium content (see above), in
Fig.~\ref{MLrelation} we also show the MLR for $\textit{Y}=0.30$,
$\textit{Z}=0.008$ (green lines). Note that this relation is slightly
more luminous than those calculated for the standard LMC elemental
composition ($\textit{Y}=0.25, \textit{Z}=0.008$).  According to the
location of the variables in the ML plane, the canonical  MLR is not
strictly satisfied, as the points are spread between the canonical and
full overshooting predictions. Even if at this stage we cannot
disentangle the role of overshooting,  mass loss and rotation in
producing the quoted excess luminosity,  at fixed mass, the
detected dispersion might indicate a combination of these different
noncanonical phenomena. Indeed, if only overshooting were efficient,
one would in principle expect the same amount of excess luminosity for
all stellar masses (within small uncertainties). Rotation
  produces similar effects as overshooting because it implies a larger
  He burning core and a brighter luminosity at fixed mass \citep[see e.g.][]{and16}
On the other  hand, if the mass loss process were efficient, this could be inferred from the predicted deviation of the best fitting stellar mass from  the value corresponding to the canonical MLR. Such a deviation is represented in Fig.~\ref{MLoss} as a function of the canonical mass (top)  and of the  pulsation period (bottom) for the CCs in our sample.  We note that the expected mass differences range from  $0\%$ to almost $\sim50\%$ and are not clearly correlated with the pulsation period or the stellar mass.

\begin{figure}
	\centering
	\begin{minipage}[c]{\columnwidth}
		\includegraphics[trim=10 10 200 450,clip,scale=0.6]{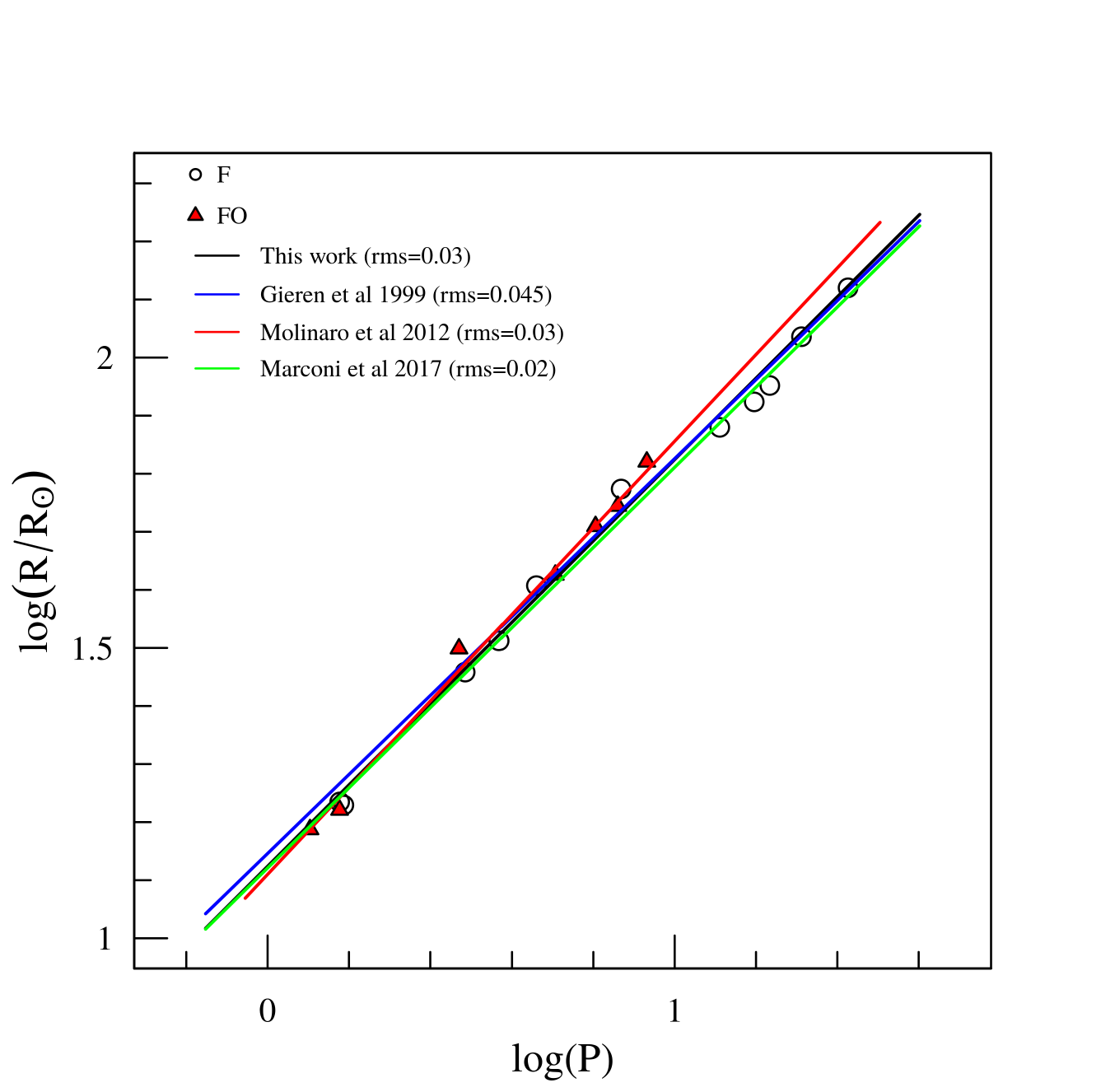}
		\captionof{figure}{The Cepheid PR relation
			obtained for the selected sample of 18
			sources.  The CCs are plotted using empty circles and filled triangles, respectively for Fs and FOs; the FO periods have been fundamentalized (see text).
			The black line is the result of a
			linear regression fit  to the data
			(see text). For comparison the PR
			relations of  \citet{gie99} (blue
			line), \citet{mol12} (red ine) and
			\citet{mar17} (green line) are also plotted.}
		\label{PRrelation}
	\end{minipage}
\end{figure}

\subsection{The Period--Mass--Radius and the Period--Radius relations}
Once we obtain the mass and the radius from the output of the non linear hydrodynamical code, we are able to correlate them with the pulsational period for each CC to investigate the PMR relation. Assuming the linearized equation introduced above, we obtain:
	\begin{multline}
	\log P = (-1.618\pm0.007) + (-0.68\pm0.02) \log (M/M_\odot) +\\ (1.72\pm 0.01)\log (R/R_\odot)
	\end{multline}  
	with a $\sigma=0.005$ dex. The values of the fitted parameters are in excellent agreement with those expected from linear theory \citep[see e.g.][]{fri72}.

If we neglect the mass dependence in the PMR relation, we can obtain the PR relation which is also well studied in the literature \citep[see e.g.][and references therein]{gie99, mol11, mol12}. The location of the $18$ investigated CCs in the (PR)  diagram is shown in Fig.~\ref{PRrelation}  where Fs are plotted using empty circles and FOs with filled triangles;  the periods of the FO pulsators have been
fundamentalized using the equation given by \citet{fea97}. A linear regression fit to the data gives us  the following PR relation:
\begin{equation}\label{PReq}
\log(R/R_{\odot})=(0.70\pm0.02)\log(P)+(1.12\pm0.01);
\end{equation}
that is displayed with a black line   in Fig.~\ref{PRrelation} with a $\sigma=0.03$ dex.
The same figure shows, for comparison, the PR relations of \citet{mol12} (red line) and \citet{gie99}, (blue line), based on two different realizations of the Baade--Wesselink method, and the relation of \citet{mar17} (green line), which was obtained using the model fitting technique for a sample of CCs in the SMC.
The PR relation of \citet{mol12}  is based on a sample of 11 CCs belonging to the young
LMC blue populous cluster NGC 1866 and 26 Galactic CCs \citep[see also][]{mol11}, while the PR relation of \citet{gie99} has been derived from a sample of both Galactic and Magellanic CCs. 
\textit{}nspection of Fig.~\ref{PRrelation} and a comparison of the coefficients of the plotted relations reveal that the PR found in this work is in agreement with those of  \citet{gie99} and \citet{mol12}. In particular, the result of \citet{gie99} predicts a shallower relation but it is in excellent agreement ($1\sigma$) with our PR, while the slope of \citet{mol12}  is steeper than that obtained  in this work, though it is consistent within $\sim2\sigma$.  Concerning the intercepts of relations, they are in excellent agreement ($1\sigma$) with that found in the current work.
The Baade--Wesselink technique is known to be dependent on the adopted value of the projection factor (p--factor), which allows to convert spectroscopically measured radial velocity into pulsational velocity \citep[see][and their references for a discussion]{gie99, mol11, mol12, gal17, ker17, nar17}. Since the radii obtained from pulsating models are not dependent on this key parameter, comparing the results from the two techniques allows us to put constraints on the p--factor. As for the cases discussed in this work, \citet{mol12} adopted a constant p--factor of 1.27, while \citet{gie99} used a period dependent value ($p=1.39 - 0.03\log P$ ) from \citet{hin86}. Since both results are consistent with the PR obtained from pulsational models, we are not able to single out one of the two choices as better.

In order to compare the PR relations by Gieren et al. and Molinaro et al., using the same projection factor,  we rescaled first the results from the former work to the constant p--factor value adopted by Molinaro et al., and then the results from the latter work to the variable p--factor used by Gieren et al.  From this procedure we can conclude that the best agreement with the PR relation in eq.\ref{PReq} is obtained by using the visual surface brightness technique from Gieren et al., but adopting a constant p--factor as in Molinaro et. al. In particular, in the quoted case we obtain the fitted PR relation $\log(R/R_\odot)=(0.699\pm0.017)\log P + (1.11\pm0.02)$, which is almost the same as that of eq.\ref{PReq}

We also compare our PR relation with that of \citet{mar17}, obtained for a sample of SMC CCs using the model fitting technique. Their fitted relation is given by $\log(R/R_\odot)=
(0.690 \pm 0.017) \log P + (1.121 \pm 0.016) $ and is fully consistent with our result, indicating that samples with different elemental compositions obey the same PR relation, in agreement with the theoretical results obtained by \citet{bon98}.
\begin{table}
	\centering
	\caption{Coefficients of the inferred Period--Radius, Period--Luminosity and Period--Wesenheit relations (in the $\textit{V}$,  $\textit{I}$, $\textit{K}_\mathrm{s}$ bands), respectively. Columns 4 to 8 represent the slope ($\alpha$) and the intecept ($\beta$) with their associated errors ($\sigma_{\alpha}$, $\sigma_{\beta}$), and the rms of the residuals around the fitted relation.}
	% latex table generated in R 3.6.1 by xtable 1.8-4 package
% Wed Aug  7 10:47:26 2019
\begin{tabular}{cccccccc}
  \hline
 & Mode & band & $\alpha$ & $\sigma_\alpha$ & $\beta$ & $\sigma_\beta$ & rms \\ 
  \hline
PR &  &  & 0.70 & 0.02 & 1.12 & 0.01 & 0.03 \\ 
   \hline\
PL & F & $V$ & $-$2.63 & 0.11 & $-$1.54 & 0.11 & 0.16 \\ 
   & FO &  & $-$3.10 & 0.16 & $-$1.95 & 0.09 & 0.13 \\ 
   & F & $I$ & $-$2.93 & 0.11 & $-$1.95 & 0.10 & 0.15 \\ 
   & FO &  & $-$3.38 & 0.15 & $-$2.38 & 0.08 & 0.12 \\ 
   & F & $K_s$ & $-$3.30 & 0.09 & $-$2.43 & 0.09 & 0.13 \\ 
   & FO &  & $-$3.70 & 0.14 & $-$2.90 & 0.07 & 0.11 \\ 
   \hline\
PW & F & $W(V,I)$ & $-$3.39 & 0.10 & $-$2.58 & 0.09 & 0.14 \\ 
   & FO &  & $-$3.83 & 0.13 & $-$3.06 & 0.07 & 0.11 \\ 
   & F & $W(V,K_s)$ & $-$3.39 & 0.09 & $-$2.55 & 0.09 & 0.13 \\ 
   & FO &  & $-$3.78 & 0.13 & $-$3.02 & 0.07 & 0.11 \\ 
   \hline
\end{tabular}
\label{tab2}	
\end{table}

\begin{figure}
	\centering
	\begin{minipage}[c]{\columnwidth}
		\centering
		\includegraphics[trim=10 10 200 350,clip,scale=0.7]{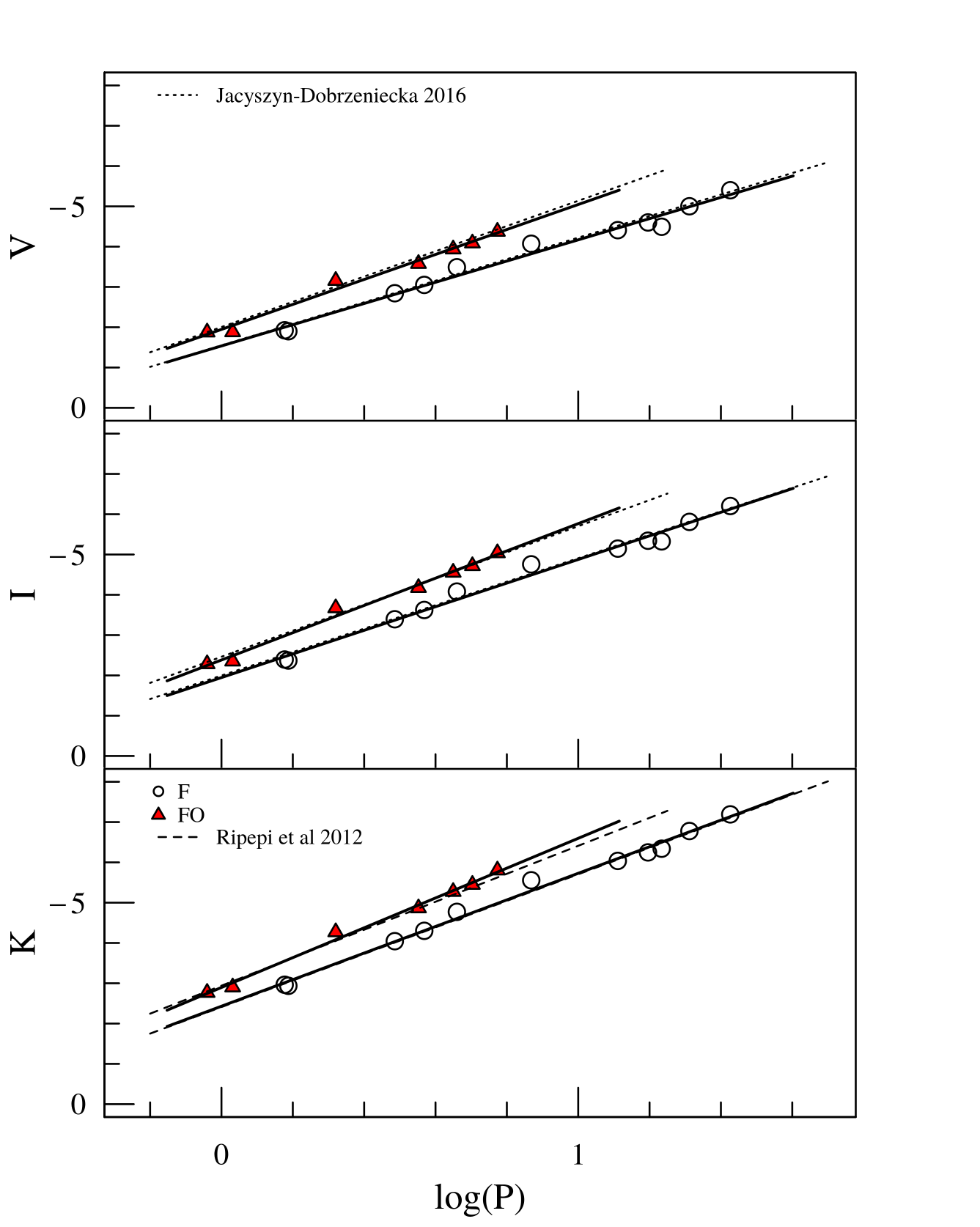}
		\captionof{figure}{Predicted PL relation in  the $V$ (top), $I$ (middle) and $K_s$ (bottom) bands based on the model fitting results for both F (empty circles) and FO (red filled triangles)
			CCs. The three panels show also  PL relations from the literature obtained, namely, by \citet{rip12b} in the $K_s$ band (dashed line)  and by \citet{jac16} in the \textit{V} and the \textit{I} bands (dotted lines).}\label{PLrelation}
	\end{minipage}
\end{figure}

\begin{figure}
	\centering
	\begin{minipage}[c]{\columnwidth }
		\includegraphics[trim=10 10 200 450,clip,scale=0.7]{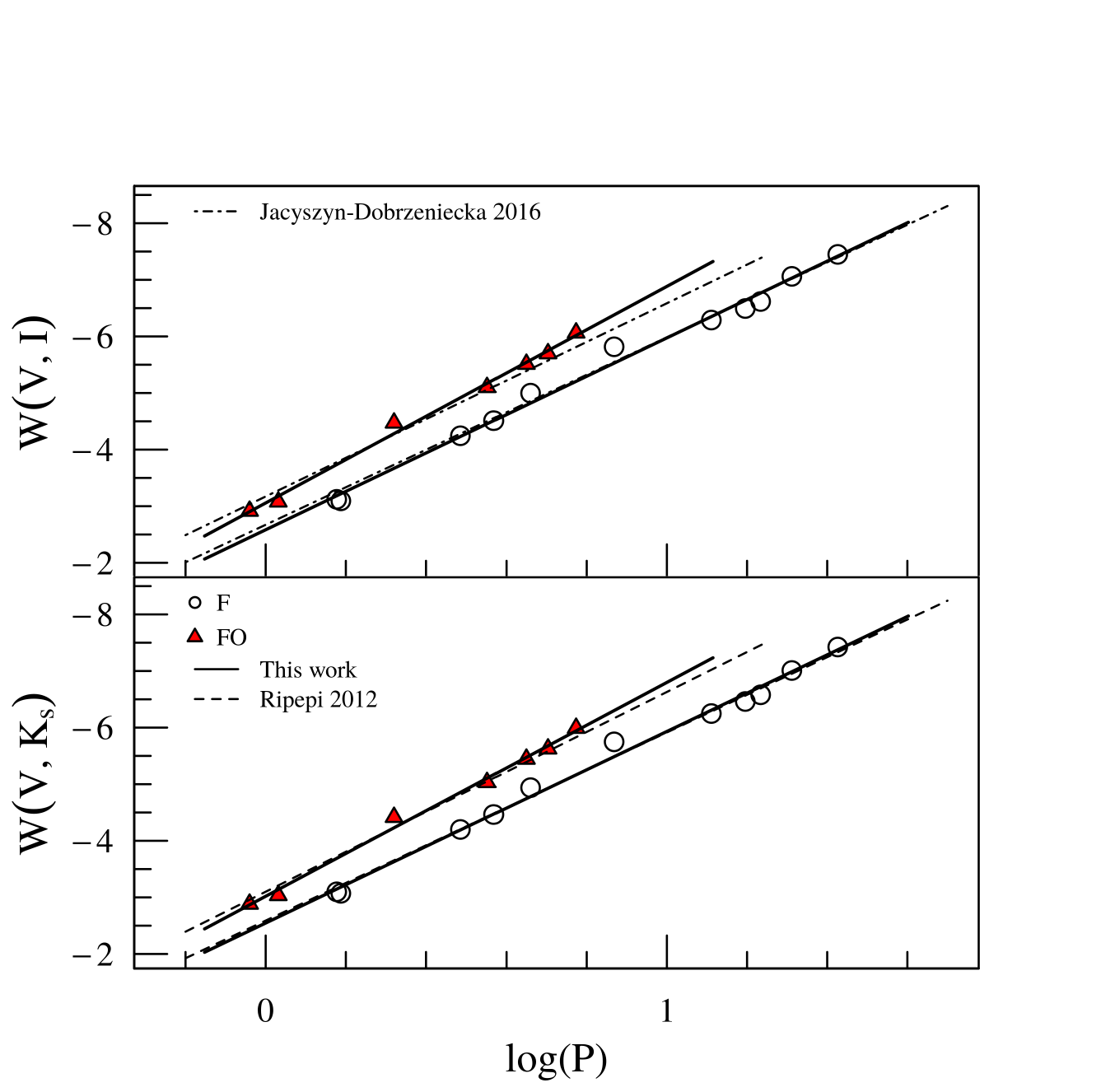}
		\captionof{figure}{Predicted PW relations in the $V,I$ (top) and $V,K_s$ (bottom) filters for both F and FO pulsators. Symbols are as in the Fig.~\ref{PLrelation}. For comparison with literature results, in the top panel we show also the relation of \citet{jac16} (dot-dashed lines), while in the bottom panel we plot the relations of \citet{rip12b} (dashed lines).}
		\label{PW}
	\end{minipage}
\end{figure}

\subsection{The PL Relation}
The mean absolute magnitudes of the best fitting models can be correlated with the corresponding periods to build multi--filter  PL relations. In Fig.~\ref{PLrelation} we show the location of both F (empty circles) and FO (red filled triangles)
best fitting models in the $\textit{V}$, $\textit{I}$ and
$\textit{K}_\mathrm{s}$ bands  versus  period.  In each panel the solid line shows the linear  regression fit to the data points whose coefficients are listed in
Table ~\ref{tab2}. To compare our theoretical relations with those obtained  by other authors, we have also plotted in the three panels the PL of \citet{jac16}, in the \textit{V} and \textit{I} bands, and of \citet{rip12b} for the $K_s$ band. 
Since the relations given by \citet{jac16} contain  apparent
magnitudes,  we have  first corrected them for absorption using
  the mean $E(B-V)$ value obtained in this work, and then shifted them
  using our best estimate of the LMC distance modulus. The comparison of the slope values provided by these authors
($\alpha^{F}_V=-2.672\pm0.006$,
$\alpha^{FO}_V=-3.133\pm0.006$, $\alpha^{F}_I=-2.911\pm0.006$
and $\alpha^{FO}_I=-3.240\pm0.006$) shows
an excellent agreement in the \textit{V} band. For the \textit{I} band the two slopes are different, but still consistent thanks to our large error. The $K_s$ band PL
relations for Fs and FOs of \citet{rip12b} contain
absolute magnitudes and consequently can be directly compared
with our results in the corresponding band. From this
comparison we found that   the coefficients of their
relation ($\alpha^F_K=-3.295\pm0.018$,
$\beta^F_K=-2.41\pm0.03$)  are in excellent agreement with
our results  for  F pulsators. As for the FO mode their zero--point ($\beta^F_K=-2.94\pm0.07$) is in excellent agreement with our value while their  slope ($\alpha^F_K=-3.471\pm0.035$) is consistent with our result within $\sim 1.5\sigma$.

Inspection of the results in Table~\ref{tab2} also suggests that
the derived rms of residuals around the PL relations are of the
same order of magnitude as that obtained for the obseravtional  relations \citep{rip12b, jac16}.

\subsection{The Wesenheit relation}
Finally, it is interesting to compare the  PW relations found in the
present work  with those adopted in the literature
\citep[e.g][]{rip12b, jac16}. To estimate the Wesenheit magnitudes for
the optical and  NIR data, we use the  following definitions:  $W(V,I) = I-1.55\times(V -I) $ and
$W(\textit{V},\textit{K}_\mathrm{s})$ = $\textit{K}_\mathrm{s} - 0.13\times(\textit{V} - \textit{K}_\mathrm{s})$, according to recent  prescriptions in the literature \citep[see e.g.][and references therein]{sos15, rip16}. The PW relations for the
investigated CCs are shown in Fig.~\ref{PW}, where the symbols
have the  same meaning as in Fig.~\ref{PLrelation}. The bottom panel shows the relation obtained by combining optical and NIR bands, and the top panel shows the same relation obtained by using only optical bands. The linear regression fits to the data are also shown (solid lines) and the coefficients are reported in Table ~\ref{tab2}. 

The PW equations in the optical bands are provided by \citet{jac16}
and are based on apparent magnitudes. In order to compare them with
the results obtained in this work, we have shifted their F and FO
relations using our best estimate of the LMC  distance modulus
$\mu=18.56$ mag. Looking at the top panel of Fig.~\ref{PW}, we
note that the relations for F Cepheids  are in excellent agreement,
being almost coincident, while the theoretical FO PW relation seems
to be steeper than the relation by \citet{jac16}. Indeed, their slope $\alpha^{FO}_{W(V,I)}=-3.414\pm0.007$ differs from our estimate by more than $3\sigma$.

The $W(V,K_s)$ relations provided by \citet{rip12b} are expressed using absolute magnitudes and consequently can be directly compared with our results. Their equation for F pulsators ($W(V,K_s)=(-3.325\pm0.014)\log P +(-2.59\pm0.03)$) is in excellent agreement with our result
(see Table~\ref{tab2}), with both slope and intercept being consistent within less than $1\sigma$. The coefficients of their FO equation ($W(V,K_s)=(-3.530\pm0.025)\log P +(-3.10\pm0.07)$), are consistent with our estimates within $\sim 1.5\sigma$.

\section{Conclusions}
We considered a sample of 11 F and 7 FO CCs in the LMC with optical photometry from the OGLE III database and NIR photometry from the VMC survey.  By assuming first approximation elemental composition typical of LMC CCs ($\textit{Y}=0.25$, $\textit{Z}=0.008$), for each selected pulsator, we built isoperiodic model sequences varying the intrinsic stellar properties (effective temperature, mass/luminosity) in order to match the period,  and the shape of the observed light curves in the $\textit{V}$, $\textit{I}$ and  $\textit{K}_\mathrm{s}$ bands. The resulting models directly
provide information on the mass, the effective temperature,
the luminosity and  in turn the individual distance of
each selected  target VMC CC. On this basis we obtained the following results:
\begin{itemize}
	\item From the inferred apparent distance moduli, adopting the extinction law by \citet{car89} we obtained an estimate of the intrinsic distance modulus for every star in our sample.  We decided to weigh these values using the best fitting $\chi^2$  to give and an estimate of the LMC distance modulus. Our procedure provides a value of $\mu_0 = 18.56$ mag with a standard deviation of $0.13$ mag, in agreement with the most recent literature values \citep{mar10,mar05, rip12a,pie13,deg14,jac16}.
	We note that our best value for the LMC distance modulus is in perfect agreement with the results of one of our previous applications \citep[ $18.53\pm 0.05$ mag][]{bon02} and the estimate by \citet{kel06}, $18.54\pm 0.018$ mag, obtained using a similar approach for a sample of bump Cepheids covering a pulsation period range centred on 10 days, an almost complementary range compared to the sample analyzed in this work.
    \item Considering the geometric correction according to the model by \citet{van01} in the estimation of the LMC distance  has no effect on the quoted results.
	\item the MLR is clearly more luminous than the evolutionary MLR that neglects overshooting, mass loss and rotation, thus suggesting a high  efficiency of at least one of these noncanonical phenomena.
	\item A PR relation in agreement with the literature results, in particular with the relation of \citet{gie99}.
	\item  Theoretical PL relations in the $\textit{V}$,
	$\textit{I}$ and  $\textit{K}_\mathrm{s}$ bands adequately reproduce the observed intrinsic scatter of the PL distribution. 
	\item Theoretical PW relations are in agreement with the  empirical LMC Wesenheit relations recently presented by \citet{rip12b}. 
\end{itemize}

We note that for one long--period Cepheid we  needed to vary the elemental composition in order to obtain a  satisfactory fit. In particular for CC $OGLE\_CEP\_LMC\_2019$ an enhanced helium  abundance $Y=0.30$ was required in order to fit the observed  curve. The possible presence of a fraction of  helium enriched CCs has been recently theoretically  investigated by \citet{car17}, following previous indications of the presence of multiple stellar populations in young LMC star clusters \citep[see e.g.][and references therein]{mil16}.

In the future we also plan to extend the application to other samples
of pulsators in order to better constrain their PL and PW relations
and to test the accuracy of the method through application to the
light curves of Galactic CCs with \textit{Gaia} parallaxes \citep{gai18}. The
latter comparison, once we have fixed the distance to the \textit{Gaia} results, will
also allow us  to put strong constraints on the predicted stellar
masses, the MLR, and, once the metallicity is precisely constrained by complementary spectroscopic data,  the helium to metal enrichment ratio.

\section*{ACKNOWLEDGEMENTS}
The authors are grateful to the referee Prof. Jan Lub for the very constructive comments that significantly improved the paper.

This work is based on observations collected at the European Organisation for Astronomical Research in the Southern Hemisphere under ESO programme 179.B-2003. We thank the CASU and the WFAU for providing calibrated data products under support of the Science and Technology Facility Council (STFC) in the UK.  
M-R.C. acknowledges funding from the European Research Council (ERC) under the European Union's Horizon 2020 research and innovation programme (grant agreement No 682115).
\bibliographystyle{plainnat}

\end{document}